\documentclass[preprint,showpacs,showkeys,preprintnumbers,amsmath,amssymb]{revtex4}

\usepackage{graphicx}
\usepackage{dcolumn}
\usepackage{bm}

\def\bea{\begin{eqnarray}}
\def\eea{\end{eqnarray}}
\def\beq{\begin{equation}}
\def\eeq{\end{equation}}
\def\bea{\begin{eqnarray}}
\def\eea{\end{eqnarray}}
\def\ve{\vert}
\def\vel{\left|}
\def\ver{\right|}
\def\nnb{\nonumber}
\def\ga{\left(}
\def\dr{\right)}

\def\rar{\rightarrow}
\def\nnb{\nonumber}

\def\ba{\begin{array}}
\def\ea{\end{array}}
\def\bea{\begin{eqnarray}}
\def\eea{\end{eqnarray}}
\def\Brll{$B\rightarrow \rho \ell^+ \ell^-$}
\def\Bpll{$B\rightarrow \pi \ell^+ \ell^-$}
\def\Bptt{$B\rightarrow \pi \tau^+ \tau^-$}
\def\Brtt{$B\rightarrow \rho \tau^+ \tau^-$}
\def\Bpee{$B\rightarrow \pi e^+ e^-$}

\begin{document}


\title{CP violation in polarized \Bpll and \Brll decays}

\author{G. Erkol}
\email{erkol@kvi.nl}
\author{J. W. Wagenaar}%
 \email{wagenaar@kvi.nl}
\affiliation{%
Theory Group, KVI \\ University of Groningen\\ Zernikelaan 25, 9747 AA Groningen\\
          The Netherlands
}%

\author{G. Turan}
\email{gsevgur@metu.edu.tr}
\affiliation{
Physics Department, Middle East Technical University\\ Inonu Bul. $06531$ Ankara\\ Turkey}%

\date{\today}

\begin{abstract}
We study the decay rate and the CP violating asymmetry of the
exclusive \Bpll and $B\rightarrow \rho \ell^+ \ell^-$ decays in
the case where one of the final leptons is polarized. We calculate
the contributions coming from the individual polarization states
in order to identify a so-called wrong sign decay, which is a
decay with a given polarization, whose width and CP asymmetry are
smaller as compared to the unpolarized one. The results are
presented for electron and tau leptons. We observe that in
particular decay channels, one can identify a wrong sign decay
which is more sensitive to new physics beyond the Standard Model.
\end{abstract}
\pacs{13.20.He, 11.30.Er, 13.88.+e}
\keywords{exclusive B meson decays, CP violation}
\maketitle

\section{Introduction}

The rare B decays, which are induced at quark level by flavour
changing neutral currents (FCNC), have received a lot of
attention, since they are very promising for investigating the
Standard Model (SM) and searching for the new physics beyond it.
Among these B-decays, the rare semileptonic ones have played a
central role for a long time, since  they offer the most direct
methods to determine the weak mixing angles and
Cabibbo-Kobayashi-Maskawa (CKM) matrix elements. These decays can
also be very useful to test the various new physics scenarios like
the two Higgs doublet models (2HDM), minimal supersymmetric
standard model (MSSM) \cite{MSSM}etc.

From experimental side, there is an impressive effort to search
for B-decays, in B-factories such as Belle, BaBar, LHC-B.  CLEO
Collaboration reports for the branching ratios (BR) of the
$B^0\rar\pi^-\ell^+\nu$ and $B^0\rar\rho^-\ell^+\nu$ decays
\cite{Cleo1} as
\bea
BR(B^0\rar\pi^-\ell^+\nu)&=&(1.8\pm 0.4\pm 0.3 \pm 0.2)\times 10^{-4}\, , \nnb\\
BR(B^0\rar\rho^-\ell^+\nu)&=&(2.57\pm 0.29_{-0.46}^{+0.33}\pm
0.41) \times 10^{-4}.
\eea
From these results, the value of the
CKM matrix element $|V_{ub}|=3.25 \pm 0.14^{+0.21}_{-0.29} \pm
0.55$ has been determined \cite{Cleo1}. Recently, the BR of the
inclusive $B\rar X_s \ell^+\ell^-$ decay has been also reported by
Belle Collaboration \cite{Belle2}; \bea BR(B\rar X_s
\ell^+\ell^-)=(6.1\pm 1.4^{+1.4}_{-1.1}) \times 10^{-6}\, , \eea
which is very close to the value predicted by the SM \cite{Ali0}. The 
experimental result from BaBar collaboration for this BR is \cite{Aubert:2003rv}:
\bea BR(B\rar X_s
\ell^+\ell^-)=(6.3\pm 1.6^{+1.8}_{-1.5}) \times 10^{-6}\, . \eea

From theoretical point of view exclusive channels are harder to
evaluate than inclusive channels, because exclusive channels
require an additional knowledge about form factors, which are used
to incorporate hadronic effects. However, the\ exclusive channels
are easier to measure. The decay channels that are induced by the
$b\rightarrow d\ell^+\ell^-$ decay at the quark level are
promising for searching the CP violation. For the B decays that
are induced by the decay $b\rightarrow s\ell^+\ell^-$, the terms
which describe virtual effects as $t\bar{t}$, $c\bar{c}$ and
$u\bar{u}$ loops are in the matrix element proportional to
$V_{tb}V_{ts}^*$, $V_{ub}V_{us}^*$ and $V_{cb}V_{cs}^*$
respectively. Because of the unitary property of the CKM matrix
and because of the fact that $V_{ub}V_{us}^*$ is small compared to
the other CKM factors, the CP violation is strongly suppressed in
these decays \cite{Aliev00,Du}. Although the BR of the B decays
induced by $b\rightarrow d\ell^+\ell^-$ are smaller, the CKM
factors $V_{tb}V_{td}^*$, $V_{ub}V_{ud}^*$ and $V_{cb}V_{cd}^*$
are all of the same order. Therefore CP violation is much more
considerable in these decays \cite{Kruger2}. In this context, the
exclusive $B_{d} \rar (\pi,\rho,\eta,\eta^\prime) \, \ell^+
\ell^-$, and $B_{d} \rar \gamma \, \ell^+ \ell^-$ decays have been
extensively studied in the SM \cite{Kruger1}-\cite{Erkol:2002nc}
and beyond \cite{Aliev4}-\cite{Choud2}.

In \cite{Babu:2003ir}, it has been observed that the unpolarized
CP asymmetry and decay width for the inclusive $b\rightarrow
d\ell^+\ell^-$ decay are comparable to the CP asymmetry and decay
width when one of the leptons is in a specific polarization state.
The CP asymmetry as well as the decay rate in the case of the
other polarization state turn out to be smaller as compared to the
unpolarized spectrum and in \cite{Babu:2003ir} this is
defined as the {\it wrong sign} polarized state. Along this line,
in \cite{Aliev:2003hw}, a similar analysis about the CP
asymmetries in  $b\rightarrow d\ell^+\ell^-$ decays has been
performed in a model independent way and it was reported that
polarized asymmetries are very sensitive to various new Wilson
coefficients. In this paper, motivated by the works in
\cite{Babu:2003ir, Aliev:2003hw}, we make a similar analysis to
the exclusive \Bpll and \Brll channels and calculate the
contributions coming from the individual polarization states in
order to identify a wrong sign decay. This feature can provide
measurements involving a new physics search.

Our paper is organized as follows. In section \ref{secexcl} we
present the effective Hamiltonian and derive the expressions for
the unpolarized and the polarized differential decay rates of
\Bpll and \Brll. The CP violating asymmetries for these decays in
the unpolarized as well as in the polarized case are calculated in
section \ref{secCP}. The numerical results and the discussions are
presented in section \ref{s3} which is followed by a conclusion
section.

\section{Exclusive \Bpll and \Brll decays}\label{secexcl}

\subsection{Effective Hamiltonian}

The leading order QCD corrected effective Hamiltonian, which is
induced by the corresponding quark level process $b \rar d \,
\ell^+ \ell^-$, is given by \cite{Wise}-\cite{Buchalla}:
\begin{eqnarray}\label{Hamiltonian} {\cal H}_{eff}  =  \frac{4
G_F \, \alpha}{\sqrt{2}} \, V_{tb} V^*_{td}\Bigg\{
\sum_{i=1}^{10}& \, \, C_i (\mu ) \, O_i(\mu)-\lambda_u
\{C_1(\mu)[O_1^u(\mu)-O_1(\mu)]\nnb\\&+C_2(\mu)[O_2^u(\mu)-O_2(\mu)]\}\Bigg\},
\end{eqnarray}
where \bea\label{CKM}
\lambda_u=\frac{V_{ub}V_{ud}^\ast}{V_{tb}V_{td}^\ast}, \eea using
the unitarity of the CKM matrix i.e.,
$V_{tb}V_{td}^\ast+V_{ub}V_{ud}^\ast=-V_{cb}V_{cd}^\ast$. The
explicit forms of the operators $O_i$ can be found in refs.
\cite{Buchalla,Wise}. In Eq.(\ref{Hamiltonian}), $C_i(\mu)$ are
the Wilson coefficients calculated at a renormalization point
$\mu$ and their evolution from the higher scale $\mu=m_W$ down to
the low-energy scale $\mu=m_b$ is described by the renormalization
group equation. For $C^{eff}_7$ this calculation is performed upto
next-to-next-to-leading logarithmic (NNLL) order in refs.\cite{Adel}-\cite{BurasA},
while $C^{eff}_{9}$ and $C_{10}$ were calculated in \cite{Bobeth}. In the context of the SM
NNLL QCD corrections to the $BR$ \cite{Bobeth}-\cite{Ghinculov1} and the forward-backward asymmetry
\cite{Ghinculov2}- \cite{Asatrian2} in $B\rightarrow X_s \ell^+ \ell^-$ are also available. For a recent
review see e.g. \cite{Hurth}. The corresponding NNLL results for $B\rightarrow X_d \ell^+ \ell^-$ are
given in \cite {Asatrian3}.

  The term that is the source of the CP
violation can be parameterized as follows:
\begin{eqnarray}\label{ksi}
C_9^{eff}=\xi_1+ \lambda_u \xi_2 ,
\end{eqnarray}
where \bea\label{Cpert1}
\xi_1&=&C_9+g(\hat{m}_c,s)(3C_1+C_2+3C_3+C_4+3C_5+C_6)\nnb\\
&&-\frac{1}{2}g(\hat{m}_d,s)(C_3+C_4)
-\frac{1}{2}g(\hat{m}_b,s)(4C_3+4C_4+3C_5+C_6)\\&&+\frac{2}{9}(3C_3+C_4+3C_5+C_6),\nnb
\eea and
\bea\label{Cpert2}\xi_2=[g(\hat{m}_c,s)-g(\hat{m}_u,s)](3C_1+C_2).\eea
In Eqs.(\ref{Cpert1}) and (\ref{Cpert2}), $s=q^2/m_B^2$ where $q$
is the momentum transfer and $\hat{m}_q=m_q/m_b$. The functions
$g(\hat{m}_q, s)$ arise from one loop contributions of the
four-quark operators $O_1-O_6$ and are given by
\begin{eqnarray}
g(\hat{m}_q, s) &=& -\frac{8}{9}\ln\hat{m}_q +
\frac{8}{27} + \frac{4}{9} y \\
& & - \frac{2}{9} (2+y) |1-y|^{1/2} \left\{\begin{array}{ll}
\left( \ln\left| \frac{\sqrt{1-y} + 1}{\sqrt{1-y} - 1}\right| -
i\pi \right), &\mbox{for } y \equiv \frac{4\hat{m}_q^2}{ s} < 1 \nonumber \\
2 \arctan \frac{1}{\sqrt{y-1}}, & \mbox{for } y \equiv \frac
{4\hat{m}_q^2}{ s} > 1 \, .
\end{array}
\right.
\end{eqnarray}
$C_9^{eff}$ term receives also contributions from long-distance
effects. The $c\bar{c}$ resonance can be parameterized by means of
a Breit-Wigner shape \cite{Ali:1991is}. It is incorporated in the
$C_9^{eff}$ term by the following replacement, \bea\label{BW}
  g(\hat{m_c},s)\rightarrow g(\hat{m_c},s)-\frac{3\pi}{\alpha^2}\kappa\sum_{V=J/\psi,\psi',...}
  \frac{m_VBr(V\rightarrow\ell^+\ell^-)\Gamma^V_{total}}{sm_B^2-m_V^2+im_V\Gamma^V_{total}}.
\eea To reproduce the correct experimental BR for $Br(B\rightarrow
J/\psi X\rightarrow X\ell\bar{\ell}) =Br(B\rightarrow J/\psi
X)Br(J/\psi\rightarrow X\ell\bar{\ell})$, the factor $\kappa$ is
taken to be $2.3$ \cite{Ali:1991is}.

Neglecting the mass of the $d$ quark, the effective short distance
Hamiltonian for the $b \rightarrow d \ell^+ \ell^-$ decay in
Eq.(\ref{Hamiltonian}) leads to the QCD corrected matrix element:
\begin{eqnarray}\label{genmatrix}
{\cal M} &=&\frac{G_{F}\alpha}{2\sqrt{2}\pi }V_{tb}V_{td}^{\ast }%
\Bigg\{C_{9}^{eff}(m_{b})~\bar{d}\gamma _{\mu }(1-\gamma _{5})b\,\bar{\ell}%
\gamma ^{\mu }\ell +C_{10}(m_{b})~\bar{d}\gamma _{\mu }(1-\gamma _{5})b\,\bar{%
\ell}\gamma ^{\mu }\gamma _{5}\ell  \nonumber \\
&-&2C_{7}^{eff}(m_{b})~\frac{m_{b}}{q^{2}}\bar{d}i\sigma _{\mu \nu
}q^{\nu }(1+\gamma _{5})b\,\bar{\ell}\gamma ^{\mu }\ell
\Bigg\}.\nonumber\\
\end{eqnarray}

\subsection{The exclusive \Bpll decay}

In this section we present the expressions for the differential
decay rate of \Bpll decay with both unpolarized and polarized
leptons. For this purpose, we need the following matrix elements,
which are written in terms of the form factors: \bea
\langle\pi(p_\pi)|\bar{d}\gamma_\mu(1-\gamma^5)b|B(p_B)\rangle&=&f^+(q^2)(p_B+p_\pi)_\mu+f^-(q^2)q_\mu\label{ff1},\\
\langle\pi(p_\pi)|\bar{d}i\sigma_{\mu\nu}q^\nu(1+\gamma^5)b|B(p_B)\rangle&=&[q^2(p_B+p_\pi)_\mu-q_\mu(m_B^2-m_\pi^2)]
f_v(q^2)\label{ff2}. \eea Here, $p_\pi$ and $p_B$ are the four
momenta of the $\pi$ and the B meson, respectively. Also $f^+$,
$f^-$ and $f_v=-\frac{f_T}{m_B+m_\pi}$ represent the relevant form factors.

From Eq. (\ref{genmatrix}), and using the matrix elements in Eqs.
(\ref{ff1}) and (\ref{ff2}), we obtain the amplitude governing the
\Bpll decay: \bea\label{MBpi}
 \cal{M}^{B\rightarrow\pi}&=&\frac{G_F\alpha}{2\sqrt{2}\pi}V_{tb}V_{td}^*
 \Bigg\{(2Ap_\pi^\mu+Bq^\mu)\bar{\ell}\gamma_\mu\ell
 +(2Gp_\pi^\mu+Dq^\mu)\bar{\ell}\gamma_\mu\gamma^5\ell\Bigg\},
\eea where \bea
 A&=&C^{eff}_9f^+-2m_BC^{eff}_7f_v,\nonumber\\
 B&=&C^{eff}_9(f^++f^-)+2\frac{m_B}{q^2}C^{eff}_7f_v(m_B^2-m_\pi^2-q^2),\nonumber\\
 G&=&C_{10}f^+,\\
 D&=&C_{10}(f^++f^-).\nonumber
\eea

Using the matrix element in Eq. (\ref{MBpi}), performing summation
over final lepton polarizations and integrating over angle
variables, the unpolarized differential decay width is obtained
as, \bea\label{dgammaunppi}
 \Big(\frac{d\Gamma^\pi}{ds}\Big)_0=\frac{G_F^2\alpha^2}{2^{10}\pi^5}|V_{tb}{V_{td}}^*|^2m_B^3~
 v~\sqrt{\lambda_\pi}~\Delta_\pi,
\eea where  \bea\label{deltapi}
 \Delta_\pi&=&\frac{1}{3}~m_B^2~\lambda_\pi(3-v^2)(|A|^2+|G|^2)+16~m_\ell^2~r_\pi~|G|^2+4~m_\ell^2~s~|D|^2\nonumber\\
 &&+~8~m_\ell^2~(1-r_\pi-s){\rm Re}[GD^*],
\eea with  $r_\pi=m_\pi^2/m_B^2$,
$\lambda_\pi=r_\pi^2+(s-1)^2-2r_\pi(s+1)$,
$v=\sqrt{1-\frac{4t^2}{s}}$ and $t=m_{\ell}/m_B$.

In order to calculate the polarized decay spectrum, we need the
final lepton polarizations. For this, one defines orthogonal unit
vectors $\vec{e}_L$, $\vec{e}_T$ and $\vec{e}_N$ such that in the
rest frame of $\ell^-$ lepton they are written as, \bea
S^{\mu}_L&\equiv&(0,\vec{e}_L)=\Bigg(0,\frac{\vec{p}_1}{|\vec{p}_1|}\Bigg)\, ,\nnb\\
S^{\mu}_N&\equiv&(0,\vec{e}_N)=\Bigg(0,\frac{\vec{k}\times\vec{p}_1}{|\vec{k}\times\vec{p}_1|}\Bigg) \, ,\nnb\\
S^{\mu}_T&\equiv&(0,\vec{e}_T)=\Bigg(0,\vec{e}^{\,-}_N\times
\vec{e}^{\,-}_L \Bigg)\,. \eea Here, $\vec{p}_1$ is the 3-vector
of $\ell^-$ lepton and $\vec{k}$ is the 3-vector of the final
meson. The longitudinal unit vector $S_L$ is boosted to the CM
frame of $\ell^{+}\ell^{-}$ by Lorentz transformation: \bea
S^{\mu}_{L,CM}& = &
\Bigg(\frac{|\vec{p}_1|}{m_\ell},\frac{E_\ell~\vec{p}_1}{m_\ell|\vec{p}_1|}\Bigg)
\, , \eea while $S_T$ and $S_N$ are not changed by the boost. The
differential decay rate of the \Bpll decay, for any spin direction
$\vec{n}$ of $\ell^-$ can be written in the following form \bea
\label{dGam1} \frac{d\Gamma^\pi(s,\vec{n})}{ds} =
\frac{1}{2}\left( \frac{d\Gamma^\pi}{ds}\right)_0 \left[1+ P^\pi_i
\vec{e}_i  \cdot \vec{n} \right]~, \eea where a sum over $i=L,T,N$
is implied. Polarization components $P^\pi_i$ in Eq. (\ref{dGam1})
are defined as \bea \label{Pi1} P^\pi_i(s) =
\frac{d\Gamma^\pi(\vec{n}=\vec{e}_i)/ds-
d\Gamma^\pi(\vec{n}=-\vec{e}_i)/ds}{
d\Gamma^\pi(\vec{n}=\vec{e}_i)/ds+
d\Gamma^\pi(\vec{n}=-\vec{e}_i)/ds}\, . \eea The resulting
expressions for the polarization asymmetries are obtained as
\bea\label{Pipi}
 P^\pi_L&=&\frac{4m_B^2}{3\Delta_\pi}~v~\lambda_\pi~{\rm
 Re}[AG^*]\, ,\nonumber\\
 P^\pi_T&=&\frac{m_B^2}{\sqrt{s}\Delta_\pi}~\sqrt{\lambda_\pi}~\pi~t~\Big({\rm Re}[AD^*]s+{\rm
 Re}[AG^*](1-r_\pi-s)\Big)\, ,\\
 P^\pi_N&=&0\, .\nonumber
\eea Our results for $P^\pi_L$ and $P^\pi_T$ agree with the ones
given in ref. \cite{Geng:1996az}. As can be seen from the explicit
expressions of $P^\pi_i$, the polarization $P^\pi_T$ is
proportional to $m_\ell$ and therefore can be significant for
$\tau$ lepton only.

\subsection{The exclusive \Brll decay}

In this section we present the expressions for the differential
decay rate for \Brll decay with both unpolarized and polarized
leptons. For this, we need the following matrix elements:
\bea\label{matrix1}
<\rho(p_\rho,\varepsilon)|\bar{d}\gamma_{\mu}(1-\gamma_5)b|B(p_B)>&=&
-\epsilon_{\mu\nu\lambda\sigma}\varepsilon^{\ast\nu}p_
{\rho}^\lambda
p_B^\sigma\frac{2V(q^2)}{m_B+m_\rho}-i\varepsilon_{\mu}^\ast
(m_B+m_\rho)A_1(q^2)\nnb\\
&+&i(p_B+p_\rho)_\mu (\varepsilon^\ast
q)\frac{A_2(q^2)}{m_B+m_\rho}+i q_\mu(\varepsilon^{\ast}
q)\frac{2m_\rho}{q^2}[A_3(q^2)\nnb\\&&-A_0(q^2)],\eea

\bea\label{matrix2}
<\rho(p_\rho,\varepsilon)|\bar{d}i\sigma_{\mu\nu}q^\nu(1+\gamma_5)b|B(p_B)>&=&
4\epsilon_{\mu\nu\lambda\sigma}
\varepsilon^{\ast\nu}p_{\rho}^\lambda
q^{\sigma}T_1(q^2)+2i[\varepsilon^\ast_\mu(m_B^2-m_\rho^2)\nnb\\
&-&(p_B+p_\rho)_\mu(\varepsilon^\ast
q)]T_2(q^2)+2i(\varepsilon^\ast q)\nnb\\&&
\Bigg(q_\mu-(p_B+p_\rho)_\mu\frac{q^2}{m_B^2-m_\rho^2}\Bigg)T_3(q^2),
\eea \bea\label{matrix3}
<\rho(p_\rho,\varepsilon)|\bar{d}(1+\gamma_5)b|B(p_B)>=\frac{-1}{m_b}2im_{\rho}(\varepsilon^\ast
q)A_0(q^2)\, , \eea
where $p_\rho$ and $\varepsilon$ denote the four momentum and
polarization vectors of the $\rho$ meson, respectively.

From Eqs. (\ref{matrix1}-\ref{matrix3}), we get the following
expression for the matrix element of   the \Brll decay:
\bea
\label{matrixBrll}\cal{M}^{B\rightarrow\rho}&=&\frac{G_F
\alpha}{2\sqrt{2}\pi}V_{tb}V_{td}^\ast \nnb \\ && \Bigg \{
\bar{\ell}\gamma_\mu(1-\gamma_5)\ell[2A\epsilon_{\mu\nu\lambda\sigma}
\varepsilon^{\ast\nu} p_\rho^\lambda p_B^\sigma +i B
\varepsilon^\ast_{\mu}-i C(p_B+p_\rho)_\mu (\varepsilon^\ast q)-i
D
(\varepsilon^\ast q)q_\mu]\nnb\\
&+& \bar{\ell}\gamma_\mu(1+ \gamma_5 )\ell[2E
\epsilon_{\mu\nu\lambda\sigma}\varepsilon^{\ast\nu} p_\rho^\lambda
 p_B^\sigma +i F \varepsilon^\ast_{\mu} -i G(\varepsilon^\ast q)(p_B+p_\rho)
-i H(\varepsilon^\ast q) q_\mu]\Bigg \} \eea where \bea
A&=&(C^{eff}_9-C_{10})\frac{V}{m_B+m_\rho}+4\frac{m_b}{q^2}C^{eff}_7 T_1,\nnb\\
B&=&(m_B+m_\rho)\Bigg( (C^{eff}_9-C_{10}) A_1+\frac{4
m_b}{q^2}(m_B^2-m^2_\rho)C^{eff}_7
T_2\Bigg),\nnb\\
C&=&(C^{eff}_9-C_{10})\frac{A_2}{m_B+m_\rho}+
4\frac{m_b}{q^2}C^{eff}_7\Bigg(T_2+\frac{q^2}{m_B^2-m_\rho^2}T_3\Bigg),\nnb\\
D&=&2(C^{eff}_9-C_{10})\frac{m_\rho}{q^2}(A_3-A_0)-4C^{eff}_7\frac{m_b}{q^2} T_3,\nnb\\
E&=&A(C_{10} \rightarrow -C_{10}),\\
F&=&B(C_{10} \rightarrow -C_{10}),\nnb\\
G&=&C(C_{10} \rightarrow -C_{10}),\nnb\\
H&=&D(C_{10} \rightarrow -C_{10}).\nnb \eea Here $A_0$, $A_1$,
$A_2$, $A_3$, $V$, $T_1$, $T_2$ and $T_3$ are the relevant form
factors.

Using the matrix element in Eq. (\ref{matrixBrll}), we find the
unpolarized differential decay width as, \bea
\Big(\frac{d\Gamma^\rho}{ds}\Big)_0&=&\frac{\alpha^2 G_F^2
m_B}{2^{12} \pi^5}|V_{tb} V^*_{td}|^2 ~v~
\sqrt{\lambda_\rho}~\Delta_{\rho}\, ,\label{unpolGam} \eea where
\bea \label{unp} \Delta_{\rho} &=& \frac{8}{3} m_B^4
\lambda_{\rho} \Big[(m_B^2 s - m_\ell^2) \ga \vel  A \ver^2 + \vel
E \ver^2 \dr + 6 m_\ell^2 \, \mbox{\rm Re}
(A E^\ast)\Big] \nnb \\
&+& 24 m_\ell^2 \, \mbox{\rm Re} (B F^\ast)+\frac{1}{r_{\rho}} m_B^4
m_\ell^2 s
 \lambda_{\rho} \,\vel D - H\ver^2 \nnb \\
&+&\frac{2}{r_{\rho}} m_B^2 m_\ell^2 \lambda_{\rho} \, \Big( \mbox{\rm
Re} [B (- D^\ast + G^\ast + H^\ast)] +
\mbox{\rm Re} [F (C^\ast + D^\ast - H^\ast)] \Big)~~~~~~~~ \nnb \\
&+& \frac{1}{2} m_\ell \, \mbox{\rm Re} [(C - G) (D^\ast -
H^\ast)]-
\frac{2}{r_{\rho}}m_B^4 m_\ell^2 \lambda_{\rho} (2+2 r_{\rho}-s)\, \mbox{\rm Re} (C G^\ast)\Big) \nnb \\
&-&\frac{2}{3r_{\rho}s} m_B^2 \lambda_{\rho} \, \Big[m_\ell^2 (2-2
r_{\rho}+s)+m_B^2 s (1-r_{\rho}-s) \Big]
\Big[\mbox{\rm Re}(B C^\ast) + \mbox{\rm Re}(F G^\ast)\Big] \nnb \\
&+&\frac{1}{3r_{\rho}s}\, \Big[2 m_\ell^2 (\lambda_{\rho}-6 r_{\rho}s)+m_B^2 s
(\lambda_{\rho}+12 r_{\rho}s) \Big]
\ga \vel B \ver^2 + \vel F \ver^2 \dr \nnb \\
&+&\frac{1}{3r_{\rho}s} m_B^4\lambda_{\rho}\, \Big( m_B^2 s
\lambda_{\rho} + m_\ell^2 [ 2 \lambda_{\rho} + 3 s (2+2 r_{\rho} - s) ]
\Big) \ga \vel C \ver^2 + \vel G \ver^2 \dr \, ,
\eea
where $\lambda_{\rho}=r^2_{\rho}+(s-1)^2-2 r_{\rho} (s+1)$ and $r_{\rho}=m^2_{\rho}/m^2_{B}$.

The polarization components are obtained in the same way as in the
previous section. The differential decay rate of the \Brll decay,
for any spin direction $\vec{n}$ of $\ell^-$ can be written in the
following form: \bea \label{dGam1rho}
\frac{d\Gamma^\rho(s,\vec{n})}{ds} = \frac{1}{2}\left(
\frac{d\Gamma^\rho}{ds}\right)_0 \left[1+ P^\rho_i \vec{e}_i \cdot
\vec{n} \right]~, \eea where a sum over $i=L,T,N$ is implied. The
resulting expressions for the polarization asymmetries are
obtained as, \bea P^\rho_L & = &
\frac{-1}{3r_{\rho}\Delta_\rho}m^2_B v \Big( 8 m^4_B s
r_{\rho}\lambda_{\rho}(\ve E\ve^2-\ve A\ve^2)-
(12 r_{\rho} s +\lambda_{\rho}) (\ve B\ve^2-\ve F\ve^2)+m^4_B \lambda^2_{\rho}(\ve G\ve^2-\ve C\ve^2)\nnb \\
& - & 2 m^2_B \lambda_{\rho}(-1+r_{\rho}+s){\rm Re}[C B^{\ast}-F G^{\ast}]\Big)~, \nnb \\
P^\rho_T & = &\frac{-1}{4 r_{\rho}\sqrt{s}\Delta_\rho}m_B m_{\ell}
\pi\sqrt{\lambda_{\rho}}\Big(m^4_B\lambda_{\rho}(r_{\rho}-1)
(\ve C\ve^2-\ve G\ve^2)+m^2_B  s (1+3 r_{\rho}-s){\rm Re}[C F^{\ast}-B G^{\ast}]\nnb \\
& + & \!\!\! 8 r_{\rho} s m^2_B {\rm Re}[(A +E)(B^{\ast}+ F^{\ast})]+
m^2_B (\lambda_{\rho}+(-1+r_{\rho}+s)(r_{\rho}-1)){\rm Re}[B C^{\ast}-F G^{\ast}]\nnb \\
& + & (-1+r_{\rho}+s)(\ve B\ve^2-\ve F\ve^2+s m^2_B {\rm Re}[(B+F)(H^{\ast}- D^{\ast})])\nnb \\
& + & m^4_B s \lambda_{\rho}{\rm Re}[(C+G)(H^{\ast}- D^{\ast})]
\Big)~ ,  \\
P^\rho_N & = & \frac{1}{4 r_{\rho}\Delta_\rho}m_B^3 m_{\ell} \pi v
\sqrt{s \lambda_{\rho}}\Big(
8{\rm Im}[E B^{\ast}+F A^{\ast}]-(1-r_{\rho}-s){\rm Im}[(B-F)( D^{\ast}-H^{\ast})]\nnb \\
& - &  \!\!\!(1+3 r_{\rho}-s){\rm Im}[(B-F)(
C^{\ast}-G^{\ast})]+m^2_B\lambda_{\rho} {\rm Im}[(C-G)( D^{\ast}-
H^{\ast})]\nnb \Big)\, . \, \eea Our results for $P^\rho_L$,
$P^\rho_N$ and $P^\rho_T$ agree with those given in \cite{Aliev1}
for the SM case. As can be seen from the explicit expressions of
$P^\rho_i$, they involve various quadratic combinations of the
Wilson coefficients and hence they are quite sensitive to the new
physics. The polarizations $P^\rho_N$ and $P^\rho_T$ are again
proportional to $m_\ell$ as in the \Bpll decay and therefore can
be significant for $\tau$ lepton only.

\section{CP Violation}\label{secCP}

\subsection{CP Violating Asymmetry in \Bpll decay}\label{CPBpll}

In \Bpll decay with unpolarized final leptons, CP violating
differential decay width asymmetry is defined as, \bea
\label{ACP1} A^\pi_{CP}(s) = \frac{(d\Gamma^\pi/ds)_0-
(d\bar{\Gamma}^\pi/ds)_0}{(d\Gamma^\pi/ds)_0-
(d\bar{\Gamma}^\pi/ds)_0 } = \frac{\Delta_{\pi}
-\bar{\Delta}_{\pi}}{\Delta_{\pi} +\bar{\Delta}_{\pi}}~, \eea
where \bea \frac{d\Gamma^\pi}{ds} = \frac{d\Gamma(B\rightarrow \pi
e^+ e^-)}{ds},~ \frac{d\bar{\Gamma}^\pi}{ds} =
\frac{d\Gamma(\bar{B}\rightarrow \bar{\pi} e^+ e^-)}{ds}~.\nnb
\eea In the SM, the Wilson coefficient  $C^{eff}_9$ is the only
one that contributes to $A_{CP}$ above since it has an imaginary
component as well as a real one, which can be parameterized as in
Eq.(\ref{ksi}). Therefore, $(d\bar{\Gamma}^\pi/ds)_0$ and
$\bar{\Delta}_\pi$ in Eq. (\ref{ACP1}) can be obtained from
$(d\Gamma^\pi/ds)_0$ and $\Delta_\pi$ by making the replacement,
\bea \left(\frac{d\bar{\Gamma}^\pi}{ds}\right)_0& = &
\left(\frac{d\Gamma^\pi}{ds}\right)_0 \mid_{\lambda_u
\rightarrow\lambda^{\ast}_u}~~~,~~~ \bar{\Delta}_{\pi} =
\Delta_{\pi}\mid_{\lambda_u
\rightarrow\lambda^{\ast}_u}\label{rep1} \, .\eea Using Eqs.
(\ref{deltapi}), (\ref{ACP1}) and (\ref{rep1}), the CP violating
asymmetry is obtained as \bea A_{CP}(s)& = & \frac{-2{\rm
Im}[\lambda_u]\Sigma_\pi(s)}{\Delta_{\pi}+2{\rm
Im}[\lambda_u]\Sigma_\pi(s)}~,\nnb \eea where \bea\label{ACP2}
\Sigma_\pi(s)=\frac{1}{3}m_B^2\lambda_\pi(3-v^2)({f^+}^2{\rm
Im}[\xi_1^*\xi_2]  -2m_B~f_vf^+~\rm{Im}[\xi_2{C_7^{eff}}^*]). \eea
When one of the leptons is polarized in \Bpll decay, CP violating
asymmetry can be defined as follows: \bea
\label{ACPP2}A_{CP}^\pi(s,\vec{n})
=\frac{d\Gamma^\pi(s,\vec{n})/ds-d\bar{\Gamma}^\pi(s,\vec{\bar{n}}=-\vec{n})/ds}
{(d\Gamma^\pi/ds)_0+(d\bar{\Gamma}^\pi/ds)_0 }~, \eea where \bea
\frac{d\Gamma^\pi(s,\vec{n})}{ds} = \frac{d\Gamma^\pi(B\rightarrow
\pi e^+ e^-(\vec{n}))}{ds}~~,~~ \frac{d\bar{\Gamma}^\pi
(s,\vec{\bar{n}})}{ds} = \frac{B\rightarrow \pi e^+
(\vec{\bar{n}})e^-)}{ds}~.\nnb \eea Here, $\vec{\bar{n}}$ is the
spin direction of the $\ell^+$ in the $\bar{B}\rightarrow
\bar{\pi} \ell^+ \ell^-$ decay. From the expression for the
polarized differential decay width for the $B\rightarrow \pi
\ell^+ \ell^-$ decay given by Eq. (\ref{dGam1}), the width for the
corresponding CP conjugated process reads, \bea \label{dGam2}
\frac{d\bar{\Gamma}^\pi(s,\vec{\bar{n}})}{ds} = \frac{1}{2}\left(
\frac{d\bar{\Gamma}^\pi}{ds}\right)_0 \left[1+ \bar{P}^\pi_i
\vec{\bar{e}}_i  \cdot \vec{\bar{n}} \right]~. \eea Since in the
CP conserving case $\bar{P}^\pi_i=-P^\pi_i$, in the general case
with the choice $\vec{\bar{e}}_i=\vec{e}_i$, $\bar{P}^\pi_i$ can
be constructed with  the replacement, \bea
\bar{P}^\pi_i=-P^\pi_i\mid_{\lambda_u \rightarrow
\lambda^\ast_u}\label{rep2}. \eea Inserting Eqs. (\ref{dGam1}) and
(\ref{dGam2}) into Eq. (\ref{ACPP2}), and setting
$\vec{\bar{n}}=\vec{n}$, the CP violating asymmetry when lepton is
polarized with $\vec{n}=\pm \vec{e}_i$ is given by, \bea
A^\pi_{CP}(s,\vec{n}=\pm \vec{e}_i)& = &\frac{1}{2}
\frac{(d\Gamma^\pi/ds)_0 \left[ 1\pm P^\pi_i \right] -
(\bar{\Gamma}^\pi/ds)_0 \left[ 1\pm \bar{P}^\pi_i  \right]}
{(d\Gamma^\pi/ds)_0+(d\bar{\Gamma}^\pi/ds)_0 }~, \nnb \eea or, by
making use of the replacements in Eq. (\ref{rep1}) and Eq.
(\ref{rep2}) we further obtain, \bea \label{ACP3} A^\pi_{CP}
(s,\vec{n}=\pm \vec{e}_i) & = & \frac{1}{2} \left\{
\frac{(d\Gamma^\pi/ds)_0- (d\bar{\Gamma}^\pi/ds)_0}
{(d\Gamma^\pi/ds)_0- (d\bar{\Gamma}^\pi/ds)_0 } \pm
\frac{(d\Gamma^\pi/ds)_0 P^\pi_i - ((d\Gamma^\pi/ds)_0 P^\pi_i)
\mid_{\lambda_u \rightarrow \lambda^\ast_u} } {(d\Gamma^\pi/ds)_0- (d\bar{\Gamma}^\pi/ds)_0 } \right\}\nnb \\
& = & \frac{1}{2} \left\{A^\pi_{CP} (s) \pm \delta A^{\pi~~i}_{CP}
(s) \right\}~. \eea The $\delta A_{CP}^i (s)$ terms in Eq.
(\ref{ACP3}) describe the modifications to the unpolarized decay
width, which can be written as, \bea \delta
A^{\pi~~i}_{CP}(s)&=&\frac{-2{\rm
Im}[\lambda_u]\delta\Sigma^i_\pi(s)}{\Delta_\pi(s)+2{\rm
Im}[\lambda_u]\Sigma_\pi(s)}\, , \eea where \bea
\delta\Sigma^L_\pi(s)&=&\frac{2}{3}~m_B^2~v~\lambda_\pi~{f^+}^2{\rm Im}[\xi_2C_{10}^*]\, ,\\
\delta\Sigma^T_\pi(s)&=&\frac{m_B^2~t~\pi~\sqrt{\lambda_\pi}}{2\sqrt{s}}~((1-r_\pi){f^+}^2+sf^+f^-)
{\rm Im}[\xi_2C_{10}^*]\, ,\\
\delta\Sigma^N_\pi(s)&=&0\, . \eea

\subsection{CP Violating Asymmetry in \Brll decay}

In $B\rightarrow \rho \ell^+ \ell^-$  decay with unpolarized final
leptons, CP violating differential decay width asymmetry is
defined as in Eq.(\ref{ACP1}) with the replacement
$\Delta_{\pi}\rar \Delta_{\rho}$ and $d\Gamma^\pi/ds\rar
d\Gamma^\rho/ds$. Using Eqs. (\ref{unp}), (\ref{ACP1}) and
(\ref{rep1}),  the CP violating asymmetry is given as,
\bea
\label{ACP2rh} A^\rho_{CP}(s)& = & \frac{-{\rm
Im}[\lambda_u]\Sigma_\rho(s)}{2\Delta_{\rho}+{\rm
Im}[\lambda_u]\Sigma_\rho(s)}~,\nnb
\eea
where
\bea \Sigma_\rho(s)
& = & \frac{4(s+2t)}{3 r_{\rho} s (1+r_{\rho})}\Bigg\{m_{B\rho}{\rm
Im}[\xi_2]\Bigg[ A_1 m_{B\rho}^2 \Big(
c_2 m^2_B (-1+r_{\rho}+s)\lambda_{\rho}+c_1 (12 r_{\rho} s +\lambda_{\rho})\Big) \nnb \\
& + & m^2_B\lambda_{\rho}\Big( 8 m^2_{\rho}r_{\rho} s V c_3+
A_2 (c_1 (-1+r_{\rho}+s)+c_2 m^2_B \lambda_{\rho})\Big)\Bigg] \nnb \\
& &\!\!\!\!\! \!\!\!\!\!-2 {\rm Im}[\xi_1\xi^{\ast}_2] \Bigg[ 2
A_1 A_2 m^2_B m^2_{B\rho}\lambda_{\rho} (-1+r_{\rho}+s)+A^2_1
m^4_{B\rho} (12 r_{\rho} s+\lambda_{\rho})\nnb\\ &+&m^4_B
\lambda_{\rho} (8 r_{\rho} s V^2+A^2_2
\lambda_{\rho})\Bigg]\Bigg\}\, , \eea with $ m_{B\rho}\equiv
m_B+m_{\rho}$ and \bea c_1 & = &
\frac{8m_bC^{eff}_7}{q^2}(m_B^2-m^2_\rho) T_2 \, \, , \, \,
c_2= 8\frac{m_b}{q^2}C^{eff}_7\Bigg(T_2+\frac{q^2}{m_B^2-m_\rho^2}T_3 \Bigg),\nnb \\
c_3 & = & \frac{8m_bC^{eff}_7}{q^2}T_1 \, \, . \eea When one of
the leptons is polarized in $B\rightarrow \rho \ell^+ \ell^-$
decay, CP violating asymmetry can be defined as follows: \bea
\label{ACP2rho} A^\rho_{CP}(s,\vec{n})
=\frac{d\Gamma^\rho(s,\vec{n})/ds-d\bar{\Gamma}^\rho(s,\vec{\bar{n}}=-\vec{n})/ds}
{(d\Gamma^\rho/ds)_0+(d\bar{\Gamma}^\rho/ds)_0 }~, \eea where \bea
\frac{d\Gamma^\rho(s,\vec{n})}{ds} = \frac{d\Gamma(B\rightarrow
\rho \ell^+ \ell^-(\vec{n}))}{ds}~~,~~ \frac{d\bar{\Gamma}^\rho
(s,\vec{\bar{n}})}{ds} = \frac{B\rightarrow \rho \ell^+
(\vec{\bar{n}})\ell^-)}{ds}~.\nnb \eea Here, $\vec{\bar{n}}$ is
the spin direction of the $\ell^+$ in the $\bar{B}\rightarrow
\bar{\rho} \ell^+ \ell^-$ decay. From the expression for the
polarized differential decay width in the $B\rightarrow \rho
\ell^+ \ell^-$ decay given by Eq. (\ref{dGam1rho}), the width for
corresponding CP conjugated process reads \bea \label{dGam2rho}
\frac{d\bar{\Gamma}^\rho(s,\vec{\bar{n}})}{ds} = \frac{1}{2}\left(
\frac{d\bar{\Gamma}^\rho}{ds}\right)_0 \left[1+ \bar{P}^\rho_i
\vec{\bar{e}}_i  \cdot \vec{\bar{n}} \right]~. \eea

Inserting Eqs. (\ref{dGam1rho}) and (\ref{dGam2rho}) into Eq.
(\ref{ACP2rho}), and setting $\vec{\bar{n}}=\vec{n}$, the CP
violating asymmetry when lepton  is polarized with $\vec{n}=\pm
\vec{e}_i$ is given by
\bea A^\rho_{CP}(s,\vec{n}=\pm \vec{e}_i)&
= &\frac{1}{2} \frac{(d\Gamma^\rho/ds)_0 \left[ 1\pm P^\rho_i
\right] - (\bar{\Gamma}^\rho/ds)_0 \left[ 1\pm \bar{P}^\rho_i
\right]} {(d\Gamma^\rho/ds)_0+(d\bar{\Gamma}^\rho/ds)_0 }~, \nnb
\eea
or, by making use of the replacements in Eq. (\ref{rep1}) and
Eq. (\ref{rep2}) with $\pi\rightarrow \rho$ we further obtain
\bea
\label{ACP3rho} A^\rho_{CP} (s,\vec{n}=\pm \vec{e}_i) & = &
\frac{1}{2} \left\{ \frac{(d\Gamma^\rho/ds)_0-
(d\bar{\Gamma}^\rho/ds)_0}{(d\Gamma^\rho/ds)_0-
(d\bar{\Gamma}^\rho/ds)_0 } \pm \frac{(d\Gamma^\rho/ds)_0 P^\rho_i
- ((d\Gamma^\rho/ds)_0 P^\rho_i)\mid_{\lambda_u \rightarrow
\lambda^\ast_u} } {(d\Gamma^\rho/ds)_0- (d\bar{\Gamma^\rho}/ds)_0
}
\right\}\nnb \\
& = & \frac{1}{2} \left\{A^\rho_{CP} (s) \pm \delta
A_{CP}^{\rho~~i} (s) \right\}~.
\eea

The $\delta A_{CP}^{\rho~i} (s)$ terms in Eq. (\ref{ACP3rho})
describe the modifications to the unpolarized decay width, which
can be written as \bea \delta A_{CP}^{\rho~~i} (s)& = &  \frac
{\mbox{\rm Im} \lambda_u ~\delta
\Sigma_\rho^i(s)}{\Delta_\rho(s)+\bar{\Delta}_\rho(s)}~, \eea
where \bea \delta\Sigma_\rho^L(s) & = & \frac{4m_B v}{3 r_{\rho}
(1+\sqrt{r_{\rho}})}{\rm Im}[\xi_2]\Bigg\{ A_1 m_{B\rho}^2 \Big(
c^{\prime}_2 m^2_B (-1+r+s)\lambda_{\rho}+c^{\prime}_1 (12 r_{\rho} s +\lambda_{\rho})\Big) \nnb \\
& + & m^2_B\lambda_{\rho}\Big( 8 m^2_{\rho}r_{\rho} s V c^{\prime}_4+
A_2 (c^{\prime}_1 (-1+r_{\rho}+s)+c^{\prime}_2 m^2_B \lambda_{\rho})\Big) \Bigg\}\, ,\nnb \\
\eea \bea \delta\Sigma_\rho^T(s) & = & \frac{m^2_B m_{\ell}\pi}{r_{\rho}
(1+\sqrt{r_{\rho}})}\frac{\sqrt{\lambda_{\rho}}}{\sqrt{s}}
\Bigg\{-A_1 m^2_{B\rho}{\rm Im}[\xi_2]\Bigg[ (-1+r_{\rho}+s)c_1^{\prime}/m^2_B  \nnb \\
& + &  (-1+r_{\rho}+s)(-1+r_{\rho})c_2^{\prime}+
s (8 r_{\rho} c_3+(-1+r_{\rho}+s) c_3^{\prime})\Bigg] \nnb \\
&+&{\rm Im}[\xi_2]\Big[8 r_{\rho} s V c_1-A_2 \lambda_{\rho}
(c_1^{\prime}+m^2_B((r_{\rho}-1)c_2^{\prime}-
s c_3^{\prime}))\Big]\nnb \\
&- & 32 m_{B\rho}  r_{\rho} s A_1 V  {\rm Im}[\xi_1\xi^{\ast}_2]\Bigg\}\,
, \eea \bea \delta\Sigma_\rho^N(s) & = & \frac{m^2_B m_{\ell}\pi
v}{2 r_{\rho} (1+\sqrt{r_{\rho}})}\sqrt{\lambda_{\rho}}\sqrt{s} {\rm Re}[\xi_2]
\Bigg\{(-A_2 D_1+A_1 D_2 m^2_{B\rho})(-1-3 r_{\rho}+s) \nnb \\
& - &  m_{B\rho} (-1+r_{\rho}+s)(A_1 D_3 m_{B\rho}-2 D_1 T_3/m_b)-8 r_{\rho}
(A_1 c_4^{\prime}m^2_{B\rho}+c_1^{\prime}V)
\Bigg\}\, , \nnb \\
\eea
where
\bea
c_1^{\prime} & = & -2m_{B\rho} A_1 C_{10} \, \, \, \, , \, \,\, \,
c_2^{\prime}=-2 A_2 C_{10}/m_{B\rho} \, , \nnb \\
c_3^{\prime}& = & -4 T_3 C_{10}/m_{b} \, \, \, \, , \, \,\, \,
c_4^{\prime} = -2 V C_{10}/m_{B\rho} \, , \eea and \bea D_1 & = &
F(C^{eff}_9\rightarrow 0)\, \, , \, \, D_2 =
G(C^{eff}_9\rightarrow 0)\, \, , \, \, D_3 =
H(C^{eff}_9\rightarrow 0)\, \, . \, \, \eea

\section{Numerical results and discussion \label{s3}}

In this section we present the numerical analysis of both the
exclusive decays \Bpll and \Brll ~for $\ell=e ,\, \tau$. We do not
present the results for $\ell=\mu$ because they are similar to the
ones for $\ell=e$. The input parameters we used in our numerical
analysis are as follows:
\begin{eqnarray}
&& m_B =5.28 \, GeV \, , \, m_b =4.8 \, GeV \, , \,m_c =1.4 \, GeV \, , \,
 m_{\tau} =1.78 \, GeV \, ,\,
 m_{e} =0.511 \, MeV, \nnb \\ &&m_{\mu} =0.106 \, GeV,\,m_{\pi}=0.14 \, GeV \, , \,m_{\rho}=0.77 \,
  GeV \, , \, m_{d}=m_{u}= m_{\pi}=0.14 \, GeV  \, ,\,\nnb\\
&&|V_{cb}|=0.044\,,\,\alpha^{-1}=129\,,\,G_f=1.17\times10^{-5}\,
{GeV}^{-2}\,,\,\tau_B=1.56 \times 10^{-12}\,s\,.
\end{eqnarray}
Using the Wolfenstein parametrization of the CKM matrix
\cite{Wolfenstein:1983yz}, $\lambda_u$ in Eq. (\ref{ksi}) can be
written as: \bea\label{lamu}
\lambda_u=\frac{\rho(1-\rho)-\eta^2-i\eta}{(1-\rho)^2+\eta^2}+O(\lambda^2).
\eea Furthermore, we use the relation \bea\label{VtbVtd}
\frac{|V_{tb} V_{td}^\ast|^2}{|V_{cb}|^2} & = &
\lambda^2[(1-\rho)^2+\eta^2]+{\cal O}(\lambda^4) \eea where
$\lambda=\sin \theta_C\simeq 0.221$ and adopt the values of the
Wolfenstein parameters as $\rho=0.25$ and $\eta=0.34$.

In order to obtain numerical results for the \Bpll and \Brll decays, we
also need the numerical values of the decay form factors. The literature on this subject
is very rich; we give some of them here. For $B\rightarrow \pi (\rho) $ form factors are calculated in
the constituent quark model \cite{Melikhov0} and using the light-cone QCD sum rules 
\cite{Ball1,Ball2}(\cite{Ball:1998kk,Ball3}). In \cite{Debbio} the results of the lattice QCD calculations
are given for the $B\rightarrow \pi , \rho$ form factors, while  
perturbative QCD approach \cite{Lu} and the so-called large energy effective theory \cite{Beneke} have 
also been employed to calculate these form factors.

\subsection{Numerical results of the exclusive \Bpll decay}

In order to obtain numerical results for the \Bpll decay, we have
made use of the results of the constituent quark model
\cite{Melikhov0}, where the form factors $f_T$ and $f_+$ can be parameterized as: \bea
f(q^2)=\frac{f(0)}{(1-q^2/M^2)[1-\sigma_1~q^2/M^2+\sigma_2~q^4/M^4]}\, . \eea 
In this model, $f_-$ is redefined as:
\bea\label{fmin}
F_0=f_+ +\frac{q^2}{(p_B+p_\pi)q}f_-\,,
\eea
and its interpolation formula is given as:
\bea
f(q^2)=\frac{f(0)}{[1-\sigma_1~q^2/M^2+\sigma_2~q^4/M^4]}\, . \eea
The parameters $f(0)$, $\sigma_1$ and $\sigma_2$ can be found in Table \ref{tabpi}. Note that for 
$f_+$ and $f_T$ a simple monopole two-parameter formula is used {\it viz.} $\sigma_2=0$.

\begin{table}[h]
\center
\begin{tabular}{|c c c c|}
\hline\hline
& $f(0)$ & $\sigma_1$ & $\sigma_2$ \\
\hline
$f_+$ & 0.29   & 0.48  &   \\
$F_0$ & 0.29  & 0.76 & 0.28  \\
$f_T$ & 0.28   & 0.48  &  \\
\hline\hline
\end{tabular}
\caption{$B\rar\pi$ transition form factors in the constituent quark model.} \label{tabpi}
\end{table}

\begin{figure}[h]
\includegraphics[scale=0.80]{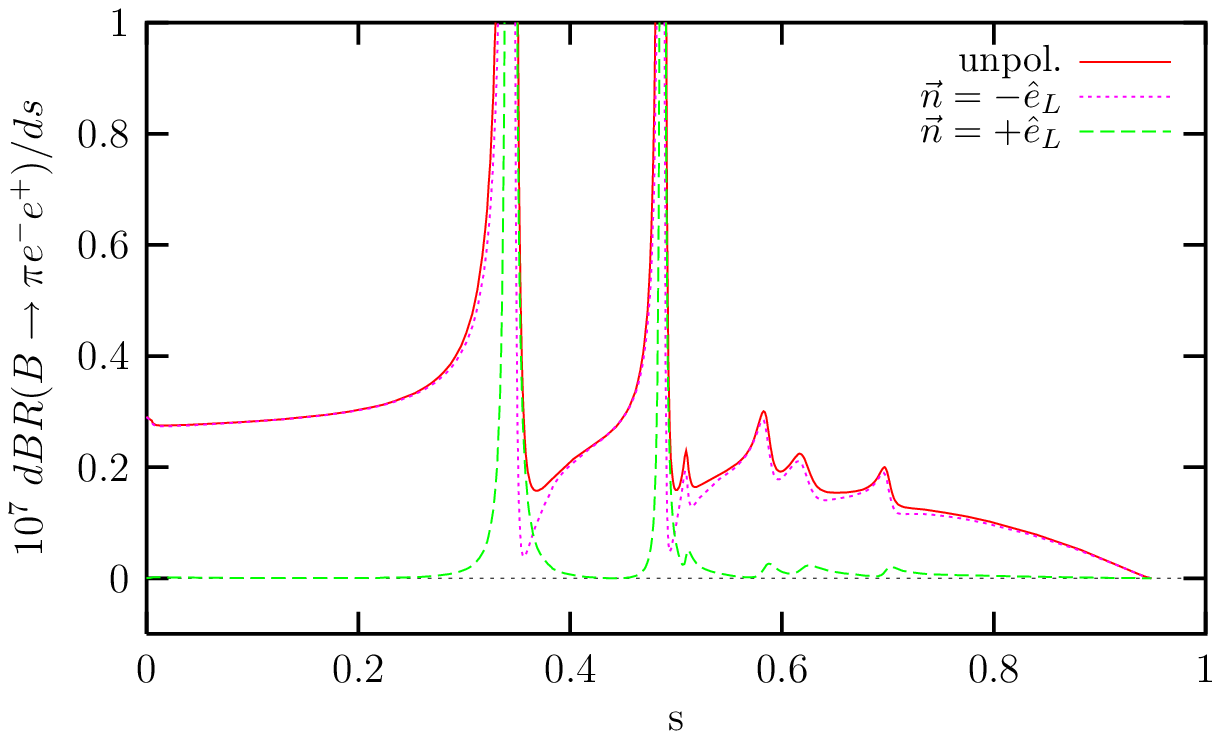}
\caption{Polarized and unpolarized differential branching ratios
for $B\rightarrow \pi e^+ e^-$ decay.} \label{fig1pi}\end{figure}

In Fig.(\ref{fig1pi}) we present our results of the differential
branching ratios (dBR/ds) of the unpolarized and longitudinally
polarized \Bpee decay. dBR/ds for $\vec{n}=-\vec{e}_L$ polarized
case is close to the one of unpolarized decay, which implies that
the decay is naturally left-handed. dBR/ds for the
$\vec{n}=+\vec{e}_L$ polarization case is far below dBR/ds for the
unpolarized one. Thus, $\vec{n}=+\vec{e}_L$ polarized \Bpee decay
corresponds to a wrong sign decay.

\begin{figure}[h]
\includegraphics[scale=0.80]{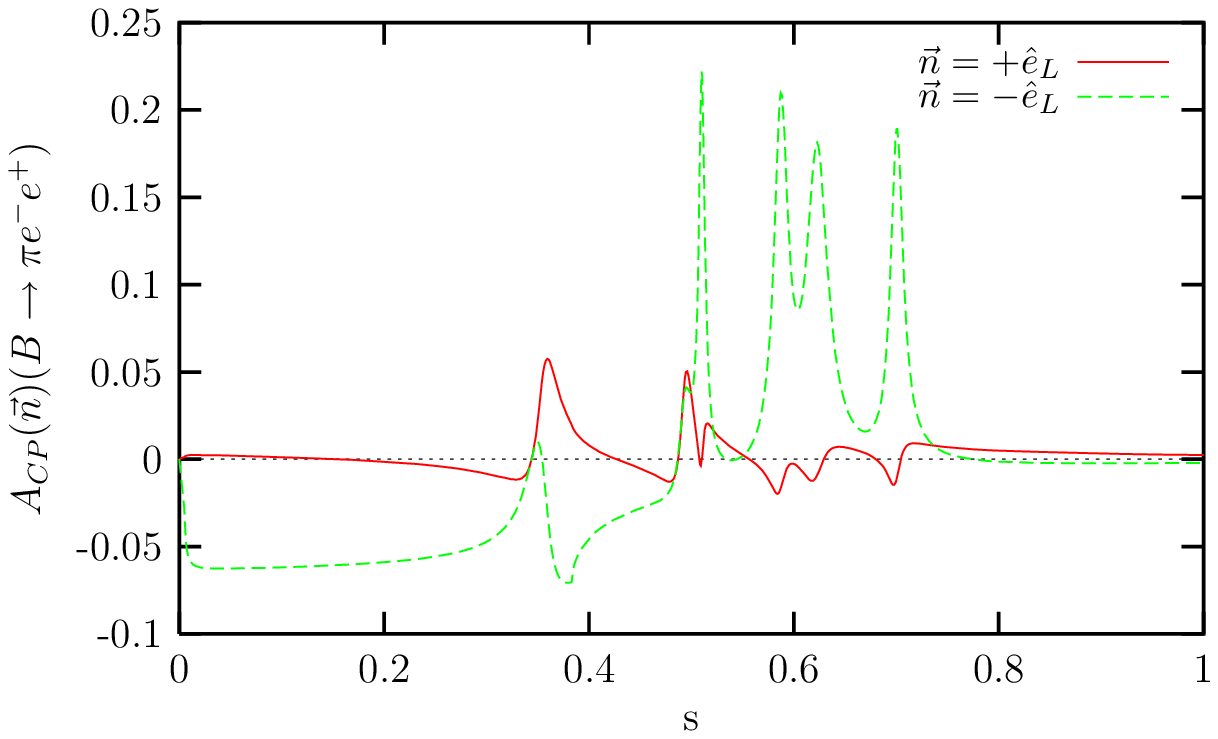}
\caption{Longitudinally polarized CP violating asymmetries for
$B\rightarrow \pi e^+ e^-$ decay.} \label{fig2pi}\end{figure}

\begin{figure}[h]
\includegraphics[scale=0.80]{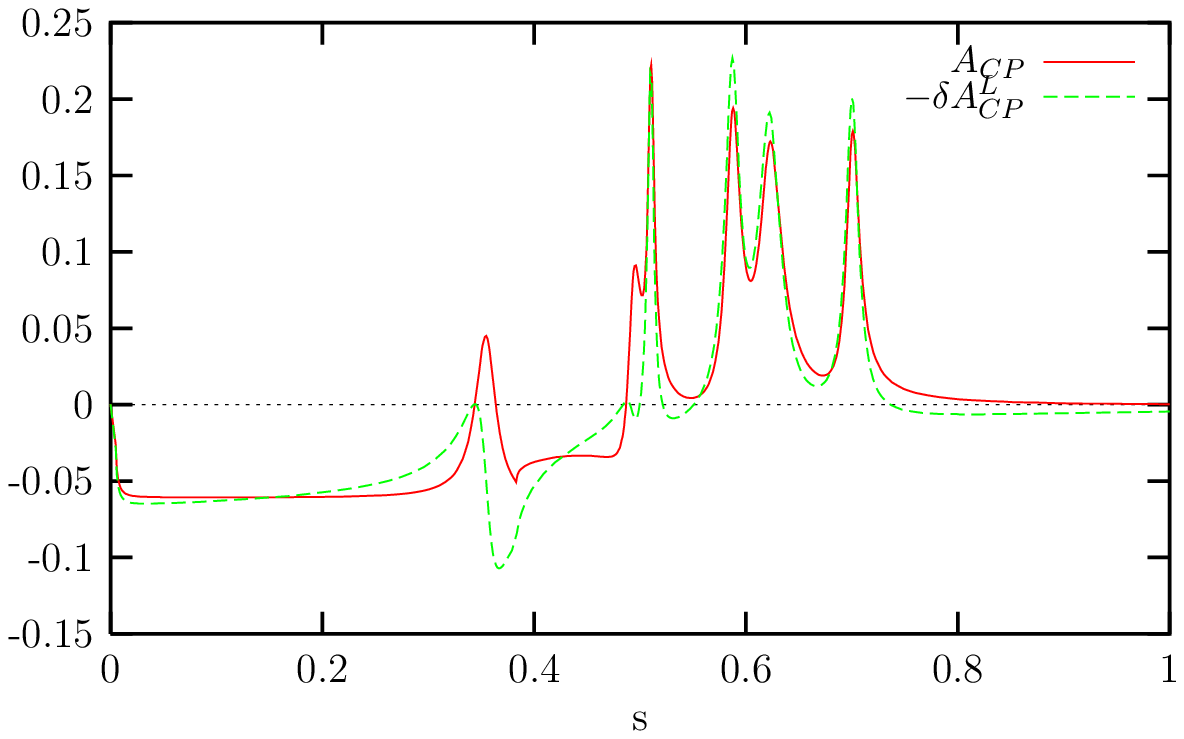}\caption{Unpolarized CP violating asymmetry and longitudinally
polarized quantity $-\delta A^L_{CP}$ for $B\rightarrow \pi e^+
e^-$ decay.} \label{fig3pi}\end{figure}

In Figs.(\ref{fig2pi}) and (\ref{fig3pi}), we plot the
longitudinally polarized asymmetries and the unpolarized CP violating asymmetry
together with $-\delta A_{CP}^L$ of the \Bpee decay, respectively.
From Fig.(\ref{fig2pi}) it can be observed that
$A_{CP}(\vec{n}=-\vec{e}_L)$ is much larger than
$A_{CP}(\vec{n}=+\vec{e}_L)$. It is also observed from
Fig.(\ref{fig3pi}) that $-\delta A_{CP}^L$ exceeds the
unpolarized $A_{CP}$ in some kinematical regions but is mostly 
comparable with it. Particularly, in the region $(2
m_{\ell}/m_B)^2 \leq s \leq ((m_{J/\psi}-0.02)/mB)^2$, which is
free of resonance contribution, we find that $\delta A^L_{CP}$ and $A_{CP}$ are 
about $6 \%$. We see also from Fig.(\ref{fig3pi}) that in the resonance region
$\delta A^L_{CP}$ can reach values up to $25 \%$.

\begin{figure}[h]
\includegraphics[scale=0.80]{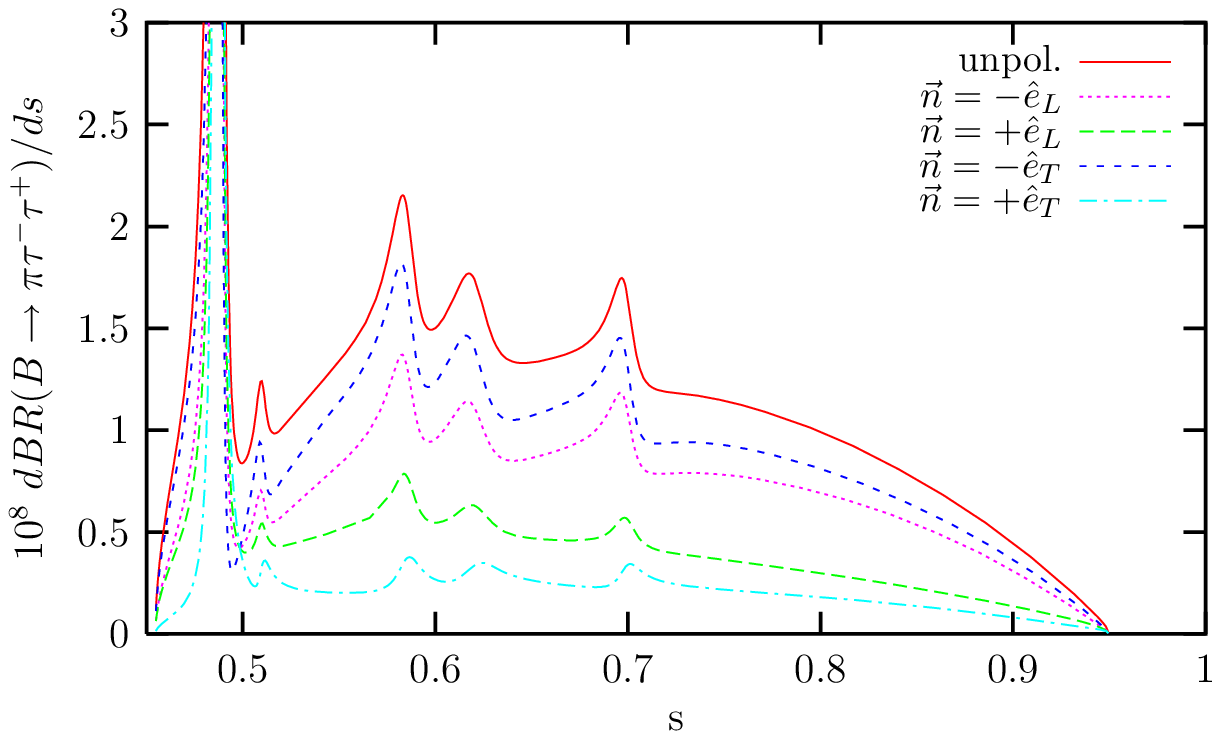}\caption{Polarized and unpolarized differential branching ratios
for $B\rightarrow \pi \tau^+ \tau^-$ decay.}
\label{fig3BRpi}\end{figure}

In Fig.(\ref{fig3BRpi}), we present dBR/ds for the decay $B
\rightarrow \pi \tau^+ \tau^-$ for unpolarized, longitudinally and
transversely polarized $\tau$ leptons. We observe that dBR/ds for
$\vec{n}=-\vec{e}_L$ and $\vec{n}=-\vec{e}_T$ are close to
unpolarized dBR/ds, while it becomes smaller for
$\vec{n}=+\vec{e}_L$ and $\vec{n}=+\vec{e}_T$. The
$\vec{n}=+\vec{e}_T$ polarization case gives a very small dBR/ds
as compared to the unpolarized decay thus can be identified as
wrong sign decay.

\begin{figure}[h]
\includegraphics[scale=0.80]{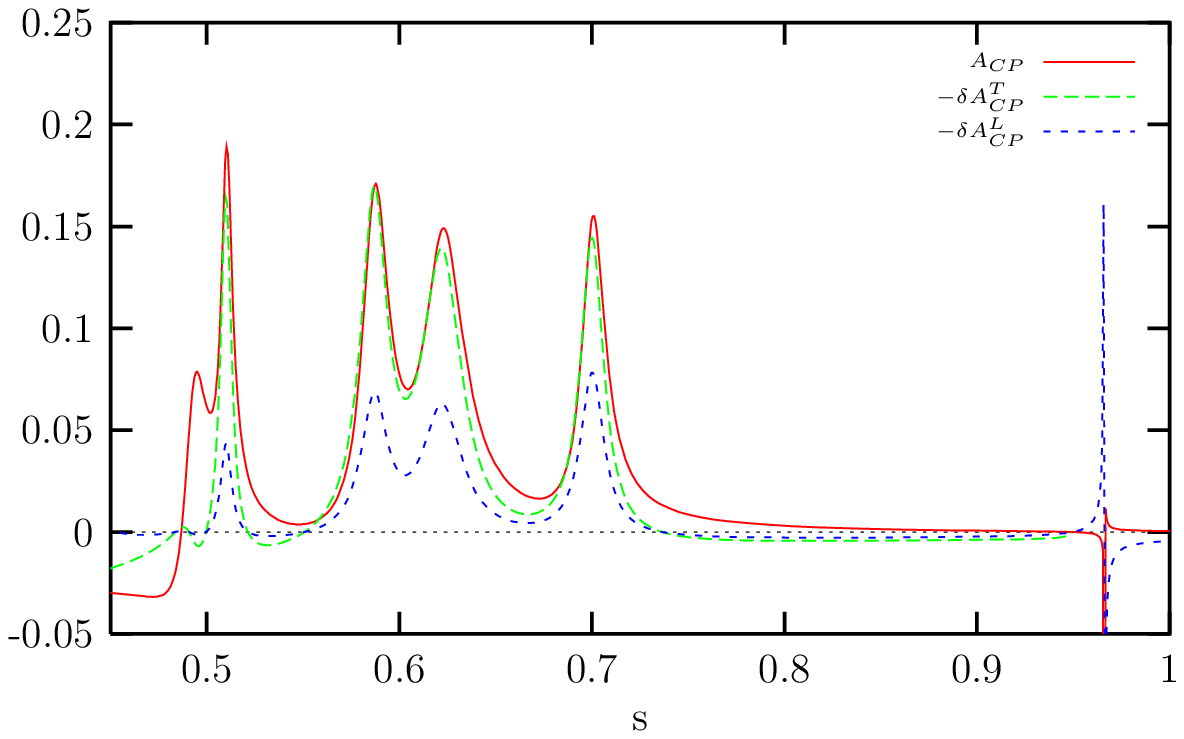}\caption{Unpolarized $A_{CP}$ and $-\delta A^i_{CP}$ with $i=L,T$ for $B\rightarrow \pi \tau^+ \tau^-$ decay.} \label{fig4pi}\end{figure}

In Fig.(\ref{fig4pi}), we plot the unpolarized $A_{CP}$ and
longitudinally and transversely polarized $-\delta A_{CP}$ of the
decay $B \rightarrow \pi \tau^+ \tau^-$. We observe that although
$\delta A^L_{CP}$ is small, $\delta A^T_{CP}$ is very close to $A_{CP}$
especially in the resonance regions. Therefore, we can conclude
that $A^L_{CP}(\vec{n}=+\vec{e}_i)\simeq
A^L_{CP}(\vec{n}=-\vec{e}_i)$. The asymmetries reach to a maximum
value of $13\%$.

\subsection{Numerical results of the exclusive \Brll decay}

In our numerical calculation for \Brll decay, we use three
parameter fit of the light-cone QCD sum rule \cite{Ball:1998kk}
which can be written in the following form:
\bea\label{formrho}
F(q^2)=\frac{F(0)}{1-a_F~q^2/m_B^2+b_F(q^2/m_B^2)^2}\, , \eea
where the values of the parameters $F(0)$, $a_F$ and $b_F$ are
given in Table (\ref{tabrho}). The form factors $A_0$ and $A_3$
can be found from the following parametrization,
\bea\label{paramet} A_0&=&A_3-\frac{T_3~q^2}{m_\rho
m_b},\nonumber\\
A_3&=&\frac{m_B+m_\rho}{2m_\rho}A_1-\frac{m_B-m_\rho}{2m_\rho}A_2.
 \eea

\begin{table}[h]
\center
\begin{tabular}{|c c c c|}\hline\hline
                  & $F(0)$ & $a_F$ & $b_F$ \\ \hline
  $A_1^{B\rar\rho}$ & $0.26\pm 0.04$ & 0.29 & -0.415 \\
  $A_2^{B\rar\rho}$ & $0.22\pm 0.03$ & 0.93 & -0.092 \\
  $V^{B\rar\rho}$ & $0.34\pm 0.05$ & 1.37 & 0.315 \\
  $T_1^{B\rar\rho}$ & $0.15\pm 0.02$ & 1.41 & 0.361 \\
  $T_2^{B\rar\rho}$ & $0.15\pm 0.02$ & 0.28 & -0.500 \\
  $T_3^{B\rar\rho}$ & $0.10\pm 0.02$ & 1.06 & -0.076 \\\hline\hline
\end{tabular}
\caption{$B\rar\rho$ transition form factors in a three-parameter
fit.} \label{tabrho}
\end{table}

\begin{figure}[h]
\includegraphics[scale=0.80]{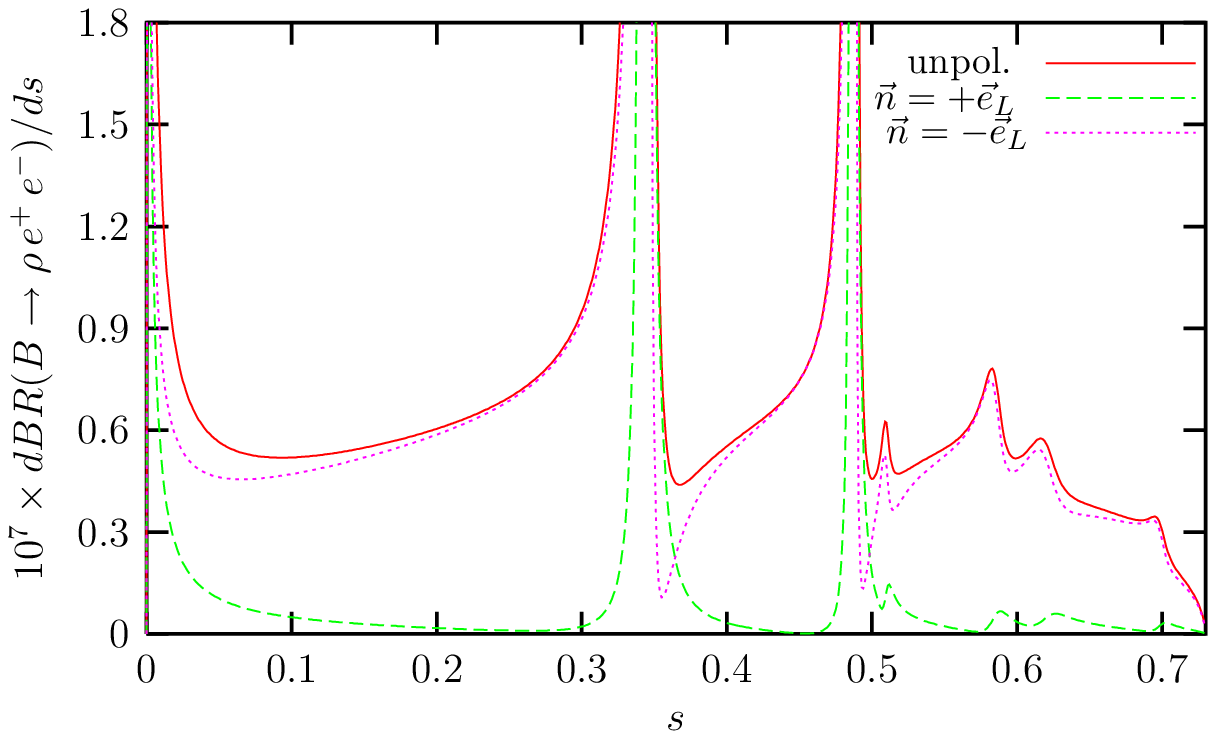}
\caption{Polarized and unpolarized differential branching ratios
for $B\rightarrow \rho e^+ e^-$ decay.} \label{fig1}\end{figure}

In Fig.(\ref{fig1}) we present dBR/ds for the decay $B \rightarrow
\rho e^+ e^-$ with unpolarized and longitudinally polarized
electrons. It can be seen from this figure that the polarized
spectrum for $\vec{n}=-\vec{e}_L$ almost coincides with
unpolarized spectrum, whereas the polarized $\vec{n}=+\vec{e}_L$
spectrum is far below the unpolarized one. So, decay is naturally
left handed in the SM.

\begin{figure}[h]
\includegraphics[scale=0.80]{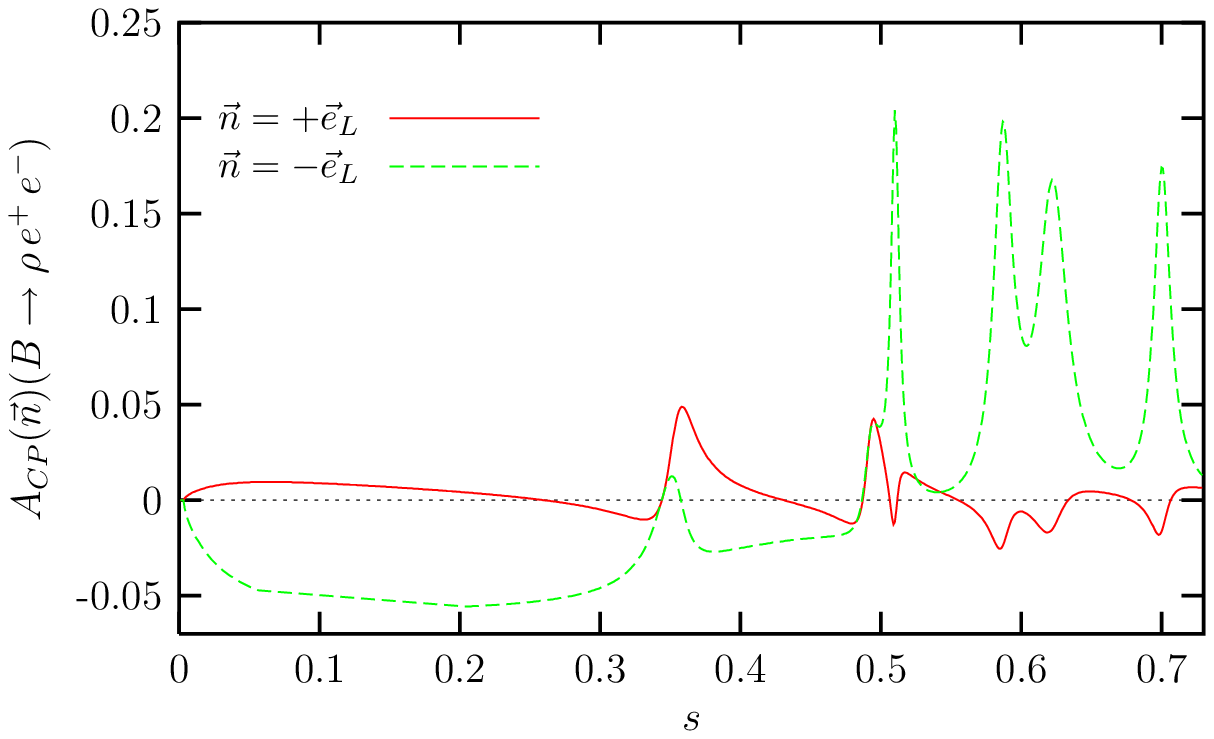}
\caption{Longitudinally polarized CP violating asymmetries for
$B\rightarrow \rho e^+ e^-$ decay.} \label{fig2}\end{figure}

\begin{figure}[h]
\includegraphics[scale=0.80]{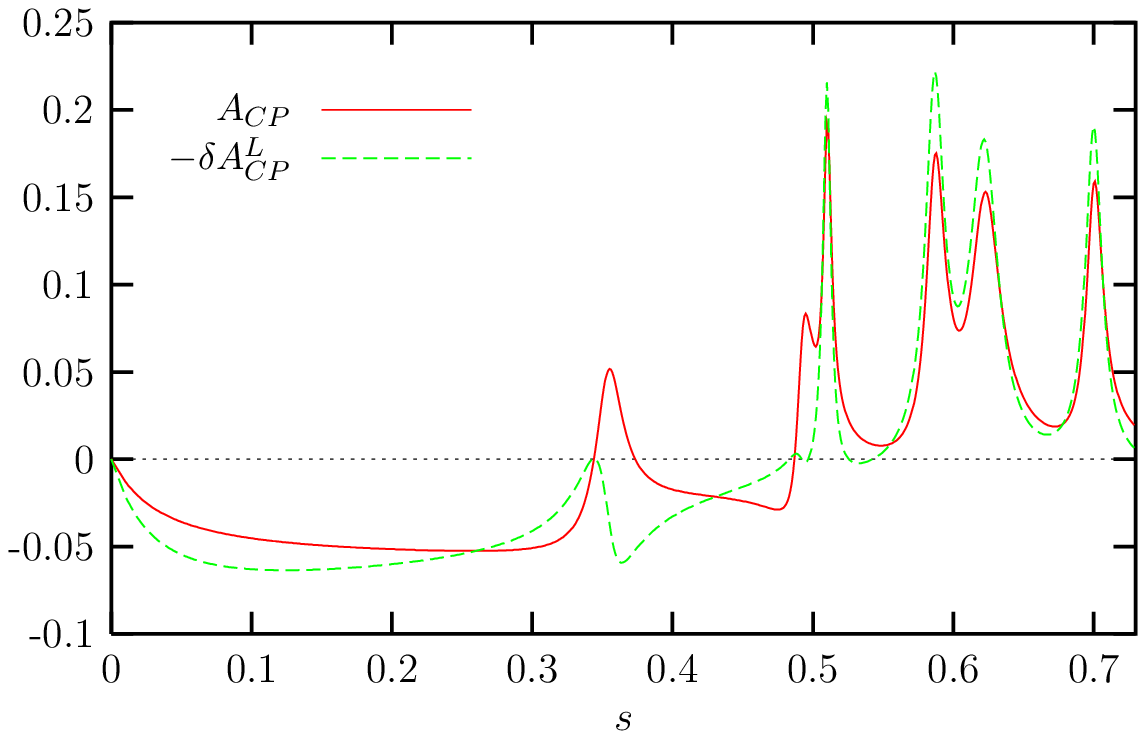}\caption{Unpolarized CP violating asymmetry and longitudinally
polarized quantity $-\delta A^L_{CP}$ for $B\rightarrow \rho e^+
e^-$ decay.} \label{fig3}\end{figure}

In Figs.(\ref{fig2}) and (\ref{fig3}) we plot the longitudinally
polarized CP violating asymmetries, $A_{CP}(\vec{n})$ with
$\vec{n}=-\vec{e}_L$ and $\vec{n}=+\vec{e}_L$, and unpolarized
$A_{CP}$ together with the polarized quantity $\delta A^L_{CP}$
for the decay $B \rightarrow \rho e^+ e^-$, respectively. As can
be seen from Fig.(\ref{fig2}), $A_{CP}(\vec{n}=-\vec{e}_L)$ is
much larger than $A_{CP}(\vec{n}=+\vec{e}_L)$. We see from
Fig.(\ref{fig3}) that polarized CP violating asymmetry $\delta
A^L_{CP}$ becomes larger than its unpolarized counterpart in some
kinematic regions. Particularly, in the region $(2 m_{\ell}/m_B)^2
\leq s \leq ((m_{J/\psi}-0.02)/mB)^2$, which is free of resonance
contribution, we find that $\delta A^L_{CP}$ is about $6 \%$,
while the unpolarized $A_{CP}$ is about $4 \%$. We see also from
Fig.(\ref{fig3}) that in the resonance region $\delta A^L_{CP}$
can reach values up to $25 \%$.

\begin{figure}[h]
\includegraphics[scale=0.80]{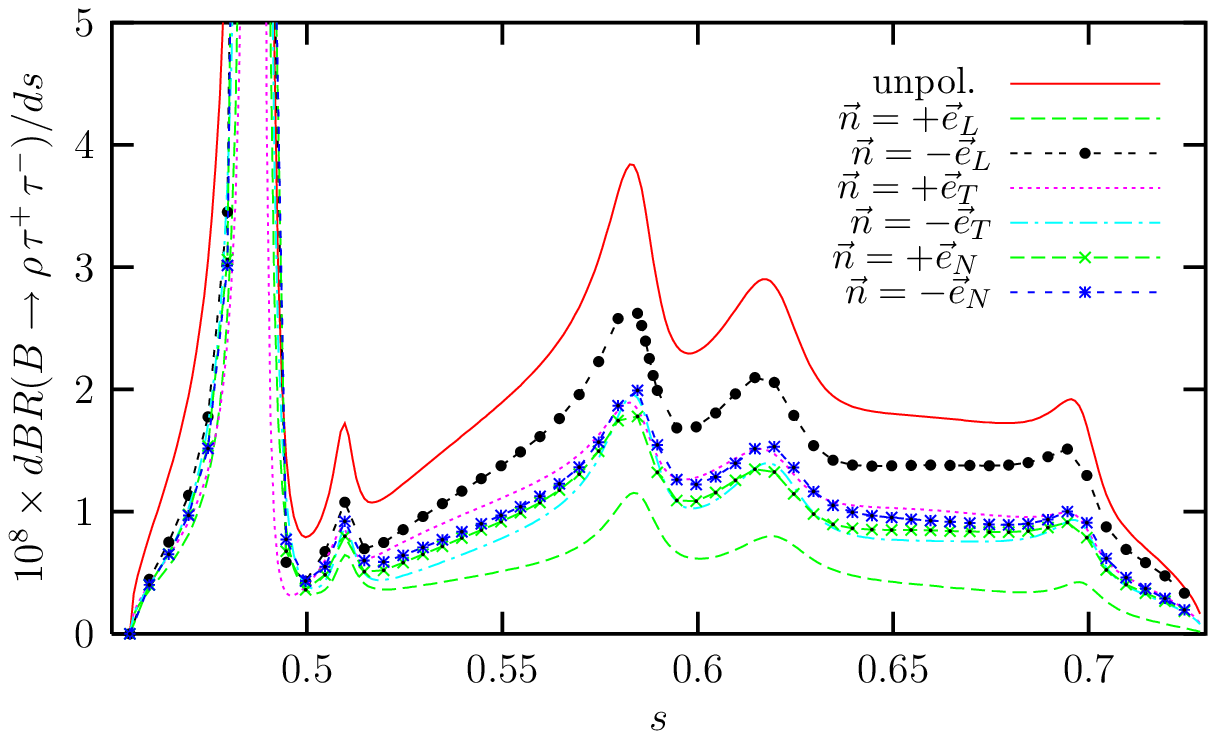}\caption{Polarized and unpolarized differential branching ratios
for $B\rightarrow \rho \tau^+ \tau^-$ decay.}
\label{fig4}\end{figure}

In Fig.(\ref{fig4}), we present the dBR/ds for the decay $B
\rightarrow \rho \tau^+ \tau^-$ for unpolarized, longitudinally,
transversely and normally polarized $\tau$ leptons. We see that
dBR/ds for $\vec{n}=+\vec{e}_N$ and $\vec{n}=-\vec{e}_N$ almost
coincide, while for $\vec{n}=\pm \vec{e}_L$, the state with
$\vec{n}=-\vec{e}_L$ is much more comparable with the unpolarized
dBR/ds with respect to the one with $\vec{n}=+\vec{e}_L$.

\begin{figure}[h]
\includegraphics[scale=0.80]{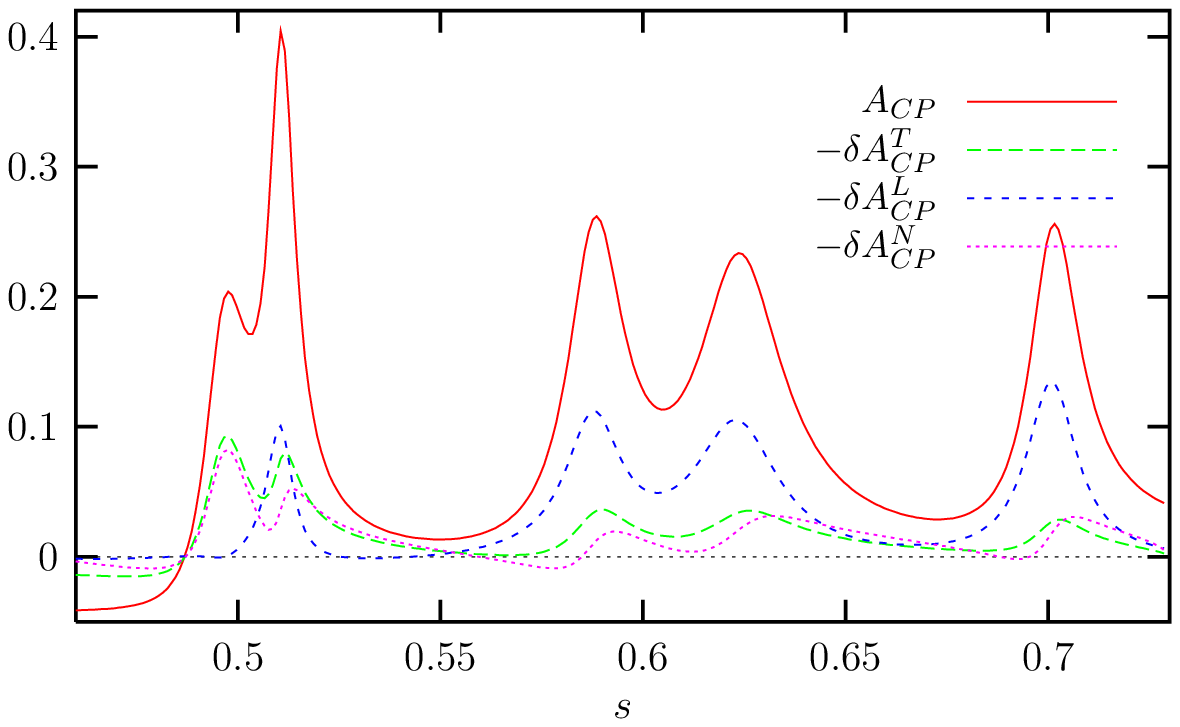}\caption{Unpolarized $A_{CP}$ and $-\delta A^i_{CP}$ with $i=L,T,N$ for $B\rightarrow \rho \tau^+ \tau^-$ decay.} \label{fig5}\end{figure}

In Fig.(\ref{fig5}), we give longitudinally, transversely and
normally polarized and unpolarized CP violating rate asymmetries
for the decay $B \rightarrow \rho \tau^+ \tau^-$. We observe that
$\delta A^T_{CP}$ and $\delta A^N_{CP}$ are both smaller than
$\delta A^L_{CP}$. Therefore, we can conclude that
$A_{CP}(\vec{n}=+\vec{e}_i)\simeq A_{CP}(\vec{n}=-\vec{e}_i)$ for
$i=T,N$, while for $i=L$ $A_{CP}(\vec{n}=+\vec{e}_L)$ is quite
small as compared to its counterpart with $\vec{n}=-\vec{e}_L$.

\section{Conclusion}

We have calculated the polarized decay rate and CP violating
asymmetries of the decays \Bpll and \Brll . For $\ell=e$ which is
in specific polarized channel $\vec{n}=-\vec{e}_L$ the decay rate
is comparable to the one of the unpolarized decay. The normal and
the transverse polarizations are proportional to the mass of the
lepton and therefore can be significant for $\tau$ lepton only.
For the \Bptt decay, $\vec{n}=\pm\vec{e}_L$ and for the \Brtt
decay $\vec{n}=\pm\vec{e}_T$ and $\vec{n}=\pm\vec{e}_N$ give
similar widths. For the rest, which are defined as the wrong sign
decays, the decay rates and the CP violating asymmetries are much
lower as compared to the unpolarized ones.

In conclusion, we studied the decay rate and the CP violating
asymmetry of the exclusive \Bpll and $B\rightarrow \rho \ell^+
\ell^-$ decays in the case where one of the final leptons is
polarized. Since the SM is naturally left-handed, the wrong sign
decays, in particular $\vec{n}=+\vec{e}_L$ polarized $B\rightarrow
(\pi,\rho) e^+ e^-$, $\vec{n}=+\vec{e}_T$ polarized $B\rightarrow
\pi \tau^+ \tau^-$ and $\vec{n}=+\vec{e}_L$ polarized
$B\rightarrow \rho \tau^+ \tau^-$ decays, are more sensitive to
new physics. Taking into account the typical branching ratios and
CP violating asymmetries, $10^{10}-10^{11}$ $B\bar{B}$ pairs are
needed for the observation of CP violation in the exclusive
channels \cite{Kruger1}, which is a challenging task for the
future hadron colliders. An unexpected large asymmetry in these
channels and the wrong sign decays would be very significant in
search for new physics beyond the SM.

\begin{acknowledgments} We would like to thank Rob Timmermans for
useful discussions. \end{acknowledgments}

\newpage




\begin{thebibliography}{100}
\bibitem{MSSM} J. L. Hewett, in L. De Porcel, C. Dunwoode (Eds.), Proc. of
the 21$^{st}$ Annual SLAC Summer Institute, SLAC-PUB-6521,1994;
J.~L.~Hewett,
\emph{hep-ph/9803370}.
%
%
\bibitem{Cleo1} CLEO Collaboration: J. P. Alexander {\it et al.}, {\it Phys. Rev. Lett.} {\bf 77} (1996) 5000; {\it Phys. Rev.} {\bf D61} (2000) 052001.
%
\bibitem{Belle2} J.~Kaneko {\it et al.}  [Belle Collaboration],
\emph{Phys.\ Rev.\ Lett.}\  {\bf 90}, 021801 (2003).
%
\bibitem{Ali0} A. Ali, E. Lunghi, C. Greub, and G. Hiller,  \emph{Phys. Rev.} D{\bf 66} (2002) 034002.
%
\bibitem{Aubert:2003rv}
B.~Aubert {\it et al.}  [BABAR Collaboration],
arXiv:hep-ex/0308016.
%
\bibitem{Aliev00} T. M. Aliev, D. A. Demir, E. Iltan and N. K. Pak,
\emph{Phys. Rev.} D{\bf 54} (1996) 851.
%
\bibitem{Du} D. S. Du and M. Z. Yang, \emph{Phys. Rev.} D{\bf 54} (1996) 882.
%
\bibitem{Kruger2} F. Kr\"{u}ger, L.M. Sehgal, \emph{Phys. Rev.} D{\bf
55} (1997) 2799.
%
\bibitem{Kruger1} F. Kr\"{u}ger, L.M. Sehgal, \emph{Phys. Rev.} D{\bf
56} (1997) 5452; D{\bf 60} (1999) 099905 (E).
%
\bibitem{Erkol} G. Erkol and G. Turan, \emph{J. Phys. } G{\bf 28} (2002) 2983.
%
\bibitem{Erkol:2002nc}
G.~Erkol and G.~Turan,
\emph{Eur.\ Phys.\ J.}\ C{\bf 28} (2003) 243.
%
\bibitem{Aliev4} T. M. Aliev and M.Savc{\i},
\emph{Phys. Rev.} D{\bf 60} (1999) 014005.
%
\bibitem{Erhan} E. O. Iltan,
\emph{Int. J. Mod. Phys.} A{\bf 14} (1999) 4365.
%
\bibitem{GurayJHEP} G. Erkol and G. Turan,
\emph{J. High Energy Phys.} {\bf 02} (2002) 015.
%
\bibitem{Choud} S. Rai Choudhury and N. Gaur,
\emph{Phys. Rev.} D{\bf 66} (2002) 094015.
%
\bibitem{Choud2} S. Rai Choudhury and N. Gaur,
\emph{arXiv:hep-ph/0207353}.
%
\bibitem{Babu:2003ir}
K.~S.~Babu, K.~R.~S.~Balaji and I.~Schienbein,
\emph{Phys.\ Rev.}\ D{\bf 68} (2003) 014021.
%
\bibitem{Aliev:2003hw}
T.~M.~Aliev, V.~Bashiry and M.~Savci,
\emph{Eur.\ Phys.\ J.}\ C{\bf 31} (2003) 511.
%
\bibitem{Wise} B. Grinstein, R. Springer, and M. Wise, \emph{
Nucl. Phys.} B{\bf 339} (1990) 269.
%
\bibitem{Buras} A. J. Buras, M. Misiak, M. M\"{u}nz, and S. Pokorski,
\emph{Nucl. Phys.} B{\bf 424} (1994) 372.
%
\bibitem{Misiak} M. Misiak,
{\it Nucl. Phys.} B{\bf 393} (1993) 23; B{\bf 439} (1993) 461 (E);
A. J. Buras and M. M\"{u}nz, \emph{Phys. Rev.} D{\bf 52} (1995)
186.
%
\bibitem{Buchalla} G. Buchalla, A. Buras, and M. Lautenbacher, \emph{
Rev. Mod. Phys.} {\bf 68} (1996) 1125.
%
\bibitem{Adel} K. Adel and Y. P. Yao, \emph{Phys. Rev.} D{\bf 49} (1994) 4945.
%
\bibitem{Greub} C. Greub and T. Hurth, \emph{ Phys. Rev.} D{\bf 56} (1997) 2934.
%
\bibitem{BurasA} A. J. Buras, A. Kwiatkowski and N. Pott,
\emph{Nucl. Phys.} B{\bf 517} (1998) 353.
%
\bibitem{Bobeth} C. Bobeth, M. Misiak and J. Urban, \emph{Nucl. Phys.} B{\bf 574} (2000) 291.
%
\bibitem{Gambino} P. Gambino, M. Gorbahn and U. Haisch, \emph{Nucl. Phys.} B{\bf 673} (2003) 238.
%
\bibitem{Asatryan1} H. H. Asatryan, H. M. Asatrian, C. Greub and M. Walker, \emph{Phys. Rev.}
D{\bf 65} (2002) 074004.
%
\bibitem{Asatryan2} H. H. Asatryan, H. M. Asatrian, C. Greub and M. Walker, \emph{Phys. Rev.}
D{\bf 66} (2002) 034009.
%
\bibitem{Ghinculov1} A. Ghinculov, T. Hurth, G. Isidori and Y. P. Yao, \emph{Eur. Phys. J.} C {\bf 33}
(2004) S288.
%
\bibitem{Ghinculov2} A. Ghinculov, T. Hurth, G. Isidori and Y. P. Yao, \emph{Nucl. Phys.} B{\bf 648} (2003)
254.
%
\bibitem{Asatrian1} H. M. Asatrian, K. Bieri, C. Greub and A. Hovhannisyan, \emph{Phys. Rev.}
D{\bf 66} (2002) 094013.
%
\bibitem{Ghinculov3} A. Ghinculov, T. Hurth, G. Isidori and Y. P. Yao,\emph{Nucl. Phys. Proc. Suppl.}
{\bf 116} (2003) 284.
%
\bibitem{Asatrian2} H. M. Asatrian, H. H. Asatryan, A. Hovhannisyan and V. Poghosyan,
 \emph{Mod. Phys. Lett.} A {\bf 19} (2004) 603.
%
\bibitem{Hurth} T. Hurth, \emph{Rev. Mod. Phys.} {\bf 75} (2003) 1159.
%
\bibitem{Asatrian3} H. M. Asatrian, K. Bieri, C. Greub and M. Walker, \emph{Phys. Rev.} D {\bf 69}
(2004) 074007.
\bibitem{Ali:1991is}
A.~Ali, T.~Mannel and T.~Morozumi,
\emph{Phys.\ Lett.}\ B{\bf 273} (1991) 505.
%
\bibitem{Geng:1996az}
C.~Q.~Geng and C.~P.~Kao,
\emph{Phys.\ Rev.}\ D{\bf 54} (1996) 5636
%
\bibitem{Aliev1} T. M. Aliev, M. K. \c{C}akmak and M. Savc{\i},{\it Nucl. Phys.} B{\bf 607} (2001) 305.
%
\bibitem{Wolfenstein:1983yz}
L.~Wolfenstein,
\emph{Phys.\ Rev.\ Lett.}\  {\bf 51} (1983) 1945.
%
\bibitem{Melikhov0} D.~Melikhov and B.~Stech, \emph{ Phys. Rev.} D {\bf 62} (2000) 014006.
%
\bibitem{Ball1} P.~Ball, \emph{JHEP} 9809 (1998) 005; {\it ibid.} 9901 (1999) 010.
%
\bibitem{Ball2} P.~Ball, \emph{arXiv:hep-ph/0308249}.
%
\bibitem{Ball:1998kk} P.~Ball and V.~M.~Braun,
\emph{Phys.\ Rev.}\ D{\bf 58}, 094016 (1998).
%
\bibitem{Ball3} P.~Ball, \emph{arXiv:hep-ph/0412079}.
%
\bibitem{Debbio} UKQCD Collaboration, L. D. Debbio {\it et al.}, \emph{Phys. Lett.} B {\bf 416} (1998)
392.

\bibitem{Lu} C. D. L\"{u} and M. Z. Yang, \emph{Eur. Phys. J.} C {\bf 28}
(2003) 515.
%
\bibitem{Beneke} M. Beneke and T. Feldmann, {\it Nucl. Phys.} B{\bf 592} (2001) 3.
%
%
%
\end{thebibliography}
\end{document}